\documentclass[10pt,aps,prl,twocolumn,longbibliography,nobibnotes]{revtex4-2}
\usepackage{color}
\usepackage{dcolumn}
\usepackage{graphicx}
\usepackage{amsmath,amsfonts,amssymb,amsthm}
\usepackage{mathtools}
\usepackage{bbold}
\usepackage{dsfont}
\usepackage{soul}
\usepackage{fancyhdr}
\usepackage{multirow}
\usepackage{bm}
\usepackage{cancel}
\usepackage{float}
\usepackage{xspace}
\usepackage{ifthen}
\usepackage{xifthen}
\usepackage{physics}

\graphicspath{{"./graphs/"}{"./scheme/"}}
\usepackage[dvipsnames,svgnames]{xcolor}
\usepackage[colorlinks,bookmarks=false,
    linkcolor=blue,
    urlcolor=blue,
    citecolor=blue
]{hyperref}

\def\op#1{\hat{#1}}
\renewcommand{\ao}[1][]{%
    \ifthenelse{\equal{#1}{}}{\ensuremath{\op{a}}}{\ensuremath{\op{#1}}}%
}
\newcommand{\co}[1][]{%
    \ifthenelse{\equal{#1}{}}{\ensuremath{{{}\op{a}^{\dagger}}}}{\ensuremath{{{}\op{#1}^{\dagger}}}}%
}
\makeatletter
\newcommand*{\transpose}{\bgroup\@transpose}
\newcommand*{\@transpose}[1][0]{\mathpalette\@@transpose{#1}\egroup}
\newcommand*{\@@transpose}[2]{\setbox0=\hbox{\m@th$#1\mkern-#2mu\intercal$}\raise\dp0\box0}
\makeatother
\usepackage{graphicx}
\newcommand{\PT}{\mathcal{PT}}

\begin{document}

\title{To $\PT$ or not to $\PT$:\\ Noise-induced escape and nonlinear-damping stabilization in a parity-time dimer}
\author{Richelle Jade L. Tuquero}
\affiliation{Department of Physics, University of Konstanz, 78464 Konstanz, Germany}
\author{Kilian Seibold}
\affiliation{Department of Physics, University of Konstanz, 78464 Konstanz, Germany}

\author{Oded Zilberberg}
\affiliation{Department of Physics, University of Konstanz, 78464 Konstanz, Germany}

\date{\today}

\begin{abstract}
Parity-time ($\mathcal{PT}$) symmetric systems exhibit long-lived excitations by balancing gain and loss in coupled resonators, driving extensive theoretical interest and diverse experimental realizations. 
Realistic physical implementations, however, inevitably introduce nonlinearities and noise. This mandates a rigorous reevaluation of their global long-time dynamics. 
In this work, we show that Hamiltonian Duffing nonlinearity restricts the $\mathcal{PT}$-unbroken phase to a finite, nonattracting region of phase space. 
Consequently, unavoidable fluctuations drive first-passage escape into runaway trajectories. This renders the linearly $\mathcal{PT}$-unbroken phase a purely transient phenomenon. 
We then recover global stochastic stability by introducing two-photon loss on the gain oscillator. 
This nonlinear damping explicitly breaks exact $\mathcal{PT}$ symmetry while supplying genuine phase-space attraction, generating a bistable regime where a low-amplitude orbit mimicking the original linear state coexists with a high-amplitude limit cycle. 
Thus, we establish a revised origin for stability in non-Hermitian experiments: the observed long-time stochastic stability is governed by inherent restoring dissipation rather than the spectral $\mathcal{PT}$ symmetry itself.
\end{abstract}

\maketitle


Parity-time ($\PT$) symmetric systems, characterized by balanced gain and loss, have attracted significant interest across classical and quantum platforms~\cite{Bender1998, Bender2007, ElGanainy2018, Ozdemir2019, bender2024pt}.
Within an ideal linear description, compensation of dissipation and amplification can yield a real spectrum and nondecaying oscillatory dynamics despite continuous exchange with the environment and regardless of initial conditions.
The real-spectrum regime terminates at an exceptional point, where eigenvalues and eigenmodes coalesce; beyond it, $\PT$ symmetry is spontaneously broken and exponentially growing and decaying components emerge~\cite{Heiss2012, Bender2007}.
Such transitions have been explored in optical waveguides~\cite{Ruter2010, Makris2008}, mechanical resonators~\cite{Bender2013, Zhang2025MechResonators}, electronic circuits~\cite{Schindler2011}, and microwave cavities~\cite{Bittner2012}.
The resulting non-Hermitian mode structure has enabled unidirectional transport~\cite{Lin2011, Feng2013, bestler2025non, schneider2025ultrafast}, enhanced sensing~\cite{Wiersig2014, Chen2017Sensitivity, Hodaei2017}, and topological mode engineering~\cite{Zhen2015, Weimann2017}.

Introducing Kerr or Duffing nonlinearities to resonator system renders the dynamics state dependent, producing bifurcations, limit cycles, and distinct behavior across phase space~\cite{Zezyulin2012, Cuevas2013, Barashenkov2014, Li2011, khedri2022fate, eichler2023classical, del2024limit, ameye2025parametric, Konotop2016, Suchkov2016}.
Therefore, one expects the $\PT$ symmetry linear framework to become insufficient once nonlinearities and fluctuations are included.
Correspondingly, the $\PT$ literature has developed along linear, nonlinear, and quantum descriptions that invoke different notions of stability with spectral $\PT$ stability, deterministic confinement, and stochastic stability manifesting, respectively, as distinct concepts.
As such, a system may be spectrally stable and only conditionally confined, while neither property alone guarantees stability under fluctuations.
At the quantum level, amplification is necessarily noisy, requiring $\PT$ symmetry to be formulated at the open-system level~\cite{Scheel2018, Huber2020} and stability to persist under continuous stochastic driving.

Early nonlinear $\PT$-dimer studies illustrate this exact separation. They identified intensity-dependent thresholds between bounded periodic motion and nonlinear runaway dynamics~\cite{Sukhorukov2010, Ramezani2010,Kevrekidis2013, Dias2014}. More recent work has emphasized the global phase-space structure involving a  coexistence of bounded and unstable dynamics, nonlinear bifurcations, and nontrivial basin geometry~\cite{Ghosh2020, Martello2023, PopDynPRR2025, NonHermitianDimer2026, Seibold2026Flow}. Stochastic investigations have likewise demonstrated the growth of ensemble-averaged intensities under gain-loss fluctuations~\cite{Konotop2014}, fluctuation-induced finite lifetimes of nonlinear oscillatory states~\cite{Nowoczyn2026Melting}, and switching between stable states or attractors~\cite{Kepesidis2016, Mukhamedyanov2025, Nowoczyn2026Escape}. In the latter, runaway dynamics were confined via dissipative processes leading to standard stochastic dynamics between stable attractors. Instead, standard Hamiltonian Duffing nonlinearity studied here can delimit a bounded region without providing relaxation. Despite this extensive catalogue of varied nonlinear instabilities, the literature persistently defaults to the linear spectral $\PT$ phase as a baseline for global stability.

\begin{figure*}[t!]
    \centering
    \includegraphics[width=\linewidth]{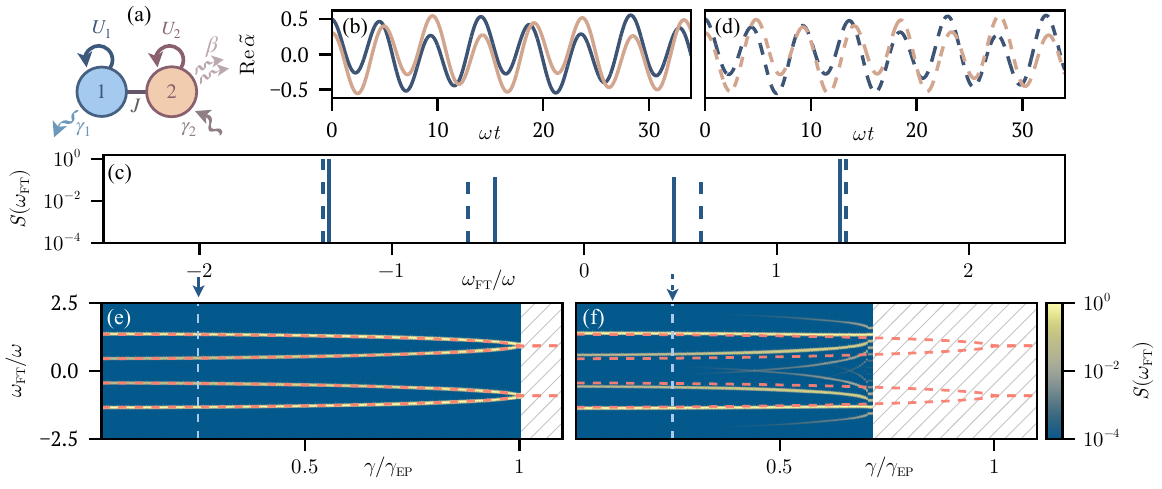}
    \caption{\textit{Linear versus nonlinear gain--loss dimer.}
    (a) Schematic of the dimer described by Eqs.~\eqref{eq:lindblad} and \eqref{eq: hamiltonian}; Hamiltonian couplings and dissipative channels are indicated by solid and wiggly lines, respectively.
    (b) Time evolution of the quadratures $\mathrm{Re}(\tilde{\alpha}_j)$ in the linear model ($U_1=U_2=0$).
    (c) Corresponding power spectral densities for the linear [solid, from (b)] and nonlinear [dashed, from (d)] dynamics, with the linear peaks occurring at $\pm\omega_\pm$ [cf.~Eq.~\eqref{eq: spectrum}].
    (d) Same as (b) for finite Duffing nonlinearities $U_1=U_2=0.1$ and $\beta=0$.
    (e) Power spectral density versus the gain--loss rate $\gamma$ for the linear model, with the analytical eigenfrequencies $\pm\omega_\pm$ overlaid as red dashed lines up to the exceptional point $\gamma_{\mathrm{EP}}$.
    (f) Same as (e) for the nonlinear model, showing amplitude-dependent frequency shifts, additional spectral components, and runaway below $\gamma_{\mathrm{EP}}$ for the chosen initial condition.
The vertical dashed lines in (e,f) indicate $\gamma=0.1$, as used in (b--d), while the hatched regions mark parameters for which the trajectories undergo runaway.
    Throughout, $\omega_1=\omega_2=0.5$, $J=0.2$, $\beta=0$, and $\gamma_{\mathrm{EP}}=2J=0.4$.
    We use the representative initial condition $\mathrm{Re}(\tilde{\alpha}_1)=0.5$, $\mathrm{Re}(\tilde{\alpha}_2)=0.3$, with vanishing imaginary parts.}
    \label{fig:sys_U}
\end{figure*}

In this work, we demonstrate that this reliance is fundamentally flawed and that the $\PT$-unbroken phase is unstable under realistic conditions. At exact gain-loss balance, the deterministic Duffing flow preserves phase-space volume, precluding asymptotically attracting sets. Consequently, the bounded sector is strictly nonattracting, leaving no restoring drift to oppose continuous fluctuations. We show that arbitrarily weak noise drives first-passage escape out of this metastable basin into runaway trajectories. There, the mean survival lifetime exhibits an effective algebraic, rather than activated exponential, scaling. The idealized linear $\PT$-unbroken phase is therefore not a long-lived physical state. To recover global stability, we introduce two-photon loss on the gain oscillator. This nonlinear damping provides genuine phase-space attraction, replacing the finite-lifetime escape with stationary stochastic dynamics between attracting limit cycles. Notwithstanding, this stabilization explicitly breaks the underlying $\PT$ symmetry. We are thus led to a strict dichotomy for interpreting existing experimental realizations. Observed stable $\PT$-like dynamics must reflect either (i) transient finite-time survival within a nonattracting bounded sector, or (ii) additional restoring mechanisms (such as uncharacterized nonlinear damping, gain saturation~\cite{Kepesidis2016}, or engineered Hamiltonian feedback~\cite{Pocklington2024}) that stabilize the system while independently breaking exact $\PT$ symmetry. Ultimately, long-time stochastic stability in nonlinear non-Hermitian systems is governed by restoring attraction, and not by the local spectrum alone.


\textit{Model}---
We consider a nonlinear dimer subject to linear gain and loss and nonlinear damping [see Fig.~\ref{fig:sys_U}(a)], described by the Lindblad master equation
\begin{equation}
    \dot{\hat{\rho}} = \frac{1}{i\hbar}[\hat{H},\,\hat{\rho}]
    + \gamma_1 \mathcal{D}[\hat{a}_1^{\phantom{\dagger}}]\hat{\rho}
    + \gamma_2 \mathcal{D}[\hat{a}_2^\dagger]\hat{\rho}
    + \beta\,\mathcal{D}[\hat{a}_2^2]\hat{\rho}\,,
    \label{eq:lindblad}
\end{equation}
where $\hat{\rho}$ is the system's density operator.
The Hamiltonian reads
\begin{align}
    \frac{\hat{H}}{\hbar} =& \sum_{j=1,2} \left[
    \omega_j\!\left(\hat{a}_j^\dagger\hat{a}_j^{\phantom{\dagger}}+\tfrac{1}{2}\right)
    + \frac{U_j}{12}(\hat{a}_j^{\phantom{\dagger}}+\hat{a}_j^\dagger)^4
    \right]\nonumber\\
    &+J(\hat{a}_1^\dagger+\hat{a}_1^{\phantom{\dagger}})(\hat{a}_2^\dagger+\hat{a}_2^{\phantom{\dagger}})\,,
    \label{eq: hamiltonian}
\end{align}
describing two oscillators ($j=1,2$) with bosonic annihilation operators $\hat{a}_j$, bare frequencies $\omega_j$, Duffing nonlinearities $U_j$, and linear inter-oscillator coupling $J$.
Each dissipator has the Lindblad form
$\mathcal{D}[\hat{o}]\hat{\rho}
=\hat{o}\hat{\rho}\hat{o}^\dagger
-\tfrac{1}{2}\{\hat{o}^\dagger\hat{o},\hat{\rho}\}$.
The three channels describe single-photon loss on oscillator~1 at rate $\gamma_1$, single-photon gain on oscillator~2 at rate $\gamma_2$, and two-photon loss on oscillator~2 at rate $\beta$, respectively.

%

For the open quantum system, we adopt the generalized $\PT$ transformation for Lindblad dynamics~\cite{Huber2020}.
Parity exchanges the two oscillators, $\mathcal{P}\hat{a}_1\mathcal{P}^{-1}=\hat{a}_2$, while the open-system $\PT$ transformation maps a jump operator $\hat{c}$ to $\mathcal{P}\hat{c}^\dagger\mathcal{P}^{-1}$.
Consequently the single-photon channels are interchanged, $\sqrt{\gamma_1}\,\hat{a}_1 \to \sqrt{\gamma_1}\,\hat{a}_2^\dagger$, so the loss channel maps onto the gain channel precisely when $\gamma_1=\gamma_2\equiv\gamma$.
Together with $\omega_1=\omega_2\equiv\omega$ and $U_1=U_2\equiv U$, this defines the $\PT$-symmetric configuration for $\beta=0$.
By contrast, the two-photon-loss channel $\hat{a}_2^2$ is mapped onto a two-photon gain channel $(\hat{a}_1^\dagger)^2$, which is absent from Eq.~\eqref{eq:lindblad}; hence $\beta\neq0$ explicitly breaks the open-system $\PT$ symmetry.
We adopt the balanced configuration throughout unless explicitly stated otherwise.
In the linear limit $U=\beta=0$, the squared eigenfrequencies of the deterministic first-moment dynamics are
\begin{equation}
    \omega_\pm^2
    =
    \omega^2-\frac{\gamma^2}{4}
    \pm\omega\sqrt{4J^2-\gamma^2}\,.
    \label{eq: spectrum}
\end{equation}
The spectrum is real for $\gamma<\gamma_{\mathrm{EP}}\equiv2J$; in the weak-coupling regime $J<\omega/2$ considered here, both $\omega_\pm^2$ remain positive throughout this interval.
At $\gamma=\gamma_{\mathrm{EP}}$, the two eigenfrequencies coalesce at an exceptional point, while for $\gamma>\gamma_{\mathrm{EP}}$ they acquire imaginary parts and the dynamics contain an exponentially growing mode.


To go beyond the deterministic nonlinear dynamics, we employ the truncated Wigner approximation (TWA), which maps the quantum master equation onto stochastic equations for complex Wigner phase-space amplitudes $\alpha_j$~\cite{Polkovnikov2010, Carusotto2013, Blakie2008, Yoneya2025}.
Within TWA, ensemble averages of these stochastic amplitudes reproduce symmetrically ordered observables to leading semiclassical order.
Introducing the rescaled amplitudes $\tilde{\alpha}_j=\alpha_j/\sqrt{\aleph}$ and retaining the stochastic contributions specified below, we obtain
\begin{align}
i\partial_t\tilde{\alpha}_j
&=
\omega_j \tilde{\alpha}_j
+J(\tilde{\alpha}_k+\tilde{\alpha}_k^*)
+\frac{\tilde U_j}{3}\left(\tilde{\alpha}_j+\tilde{\alpha}_j^*\right)^3
\label{eq:TWA}\\
&\quad
+\frac{i}{2}s_j\gamma_j\tilde{\alpha}_j
-i\delta_{j2}\tilde{\beta}\,\tilde{\alpha}_j
\left(|\tilde{\alpha}_j|^2-\frac{1}{\aleph}\right)
+\tilde{\xi}_j(t)\,,\nonumber
\end{align}
where $k=3-j$, $s_j=(-1)^j$, $\tilde U_j=\aleph U_j$, and $\tilde{\beta}=\aleph\beta$.
Here, $\aleph$ is a dimensionless quantum-to-classical scaling parameter that sets the characteristic occupation, $|\alpha_j|^2\propto\aleph$.
At fixed $\tilde U_j$ and $\tilde{\beta}$, the leading deterministic mean-field dynamics remain unchanged, while the noise amplitude in the rescaled equations scales as $\aleph^{-1/2}$, such that $\aleph\to\infty$ defines a controlled semiclassical limit~\cite{Nowoczyn2026Melting,Nowoczyn2026Escape,Casteels2017}.
For the single-photon gain and loss channels, the additive noises have zero mean,
$\langle\tilde{\xi}_j(t)\rangle=0$, and correlations
\begin{align}
\langle\tilde{\xi}_j^*(t)\tilde{\xi}_k(t')\rangle
=
\frac{\gamma_j}{2\aleph}\,
\delta_{jk}\delta(t-t')\,;
\quad \langle\tilde{\xi}_j(t)\tilde{\xi}_k(t')\rangle
=0\,.
\label{eq: scaled noise}
\end{align}
For $\beta\neq 0$, the two-photon-loss channel additionally generates multiplicative noise which we neglect in our simulations~\cite{Supp}; we do so to isolate the effect of the nonlinear dissipative drift while retaining the stochastic fluctuations associated with the linear gain and loss channels.
In the semiclassical limit $\aleph\rightarrow\infty$, the stochastic terms vanish and Eq.~\eqref{eq:TWA} reduces to the deterministic mean-field equations.
These equations possess the fixed point $\tilde{\alpha}_1=\tilde{\alpha}_2=0$, whose linearization reproduces the linear $\PT$ spectrum, cf.~Eq.~\eqref{eq: spectrum}.
This local stability criterion, however, does not determine the finite-amplitude dynamics.


To expose the limitations of the local linear description, we compare the evolution of finite-amplitude initial conditions in the linear and nonlinear dimers.
In the linear $\PT$-unbroken regime, the dynamics remain bounded and oscillatory [Fig.~\ref{fig:sys_U}(b)], with four power-spectral-density peaks at the normal-mode frequencies $\pm\omega_{\pm}$ [Fig.~\ref{fig:sys_U}(c)].
Repeating the same initialization with finite Duffing nonlinearities ($U_1=U_2\neq0$; $\beta=0$) yields similarly bounded oscillations over the displayed time window [Fig.~\ref{fig:sys_U}(d)], but with shifted frequencies and additional spectral components [Fig.~\ref{fig:sys_U}(c)].
Sweeping the gain-loss rate $\gamma$ makes the distinction explicit.
The linear spectrum follows the analytical eigenfrequencies $\omega_{\pm}$ up to their coalescence at $\gamma_{\mathrm{EP}}$ [Fig.~\ref{fig:sys_U}(e)], whereas the nonlinear response becomes amplitude dependent: the dominant frequencies shift, additional spectral components emerge, and, for the chosen initial condition, runaway occurs already at $\gamma<\gamma_{\mathrm{EP}}$ [Fig.~\ref{fig:sys_U}(f)].
Consistent with earlier studies of nonlinear $\PT$ dimers~\cite{Sukhorukov2010,Ramezani2010,Kevrekidis2013}, the linear exceptional point therefore characterizes only the local dynamics near the origin and does not delimit the finite-amplitude region of bounded motion.

\begin{figure}[tbp]
    \centering
    \includegraphics[width=1\columnwidth]{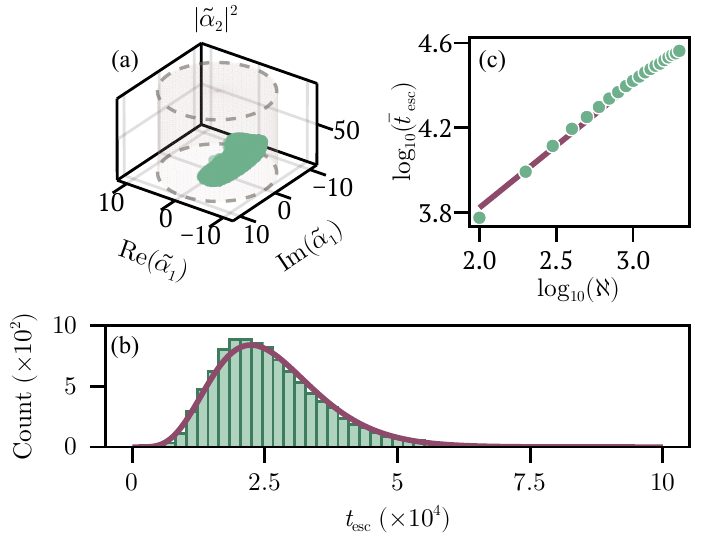}
    \caption{\textit{Deterministic confinement and noise-induced first-passage escape.}
(a) Projection of the initial-condition space onto $(\operatorname{Re}(\tilde{\alpha}_1), \operatorname{Im}(\tilde{\alpha}_1), |\tilde{\alpha}_2|^2)$, showing initial conditions that yield bounded deterministic trajectories (green), together with the escape threshold $|\tilde{\alpha}_j|^2=10^2$ (cylinder).
    (b) Distribution of first-passage escape times, $t_\mathrm{esc}$, for $N_\mathrm{traj}=10^4$ realizations at $\aleph=1000$, where $t_\mathrm{esc}$ is the first time either oscillator exceeds the escape threshold.
    The solid red line shows an empirical maximum-likelihood Gamma fit, with mean $\bar{t}_\mathrm{esc}=2.6\times10^{4}$ and relative width $\Delta t_\mathrm{esc}/\bar{t}_\mathrm{esc}=0.39$.
    (c) Mean escape time $\bar{t}_\mathrm{esc}$ versus $\aleph$, together with an effective power-law fit $\bar{t}_\mathrm{esc}\propto\aleph^{0.59}$ over the explored range (solid line).
    Throughout, system parameters match Fig.~\ref{fig:sys_U}(d).}
    \label{fig:traj_bound}
\end{figure}


\begin{figure*}[htbp]
\centering
\includegraphics[width=1\linewidth]{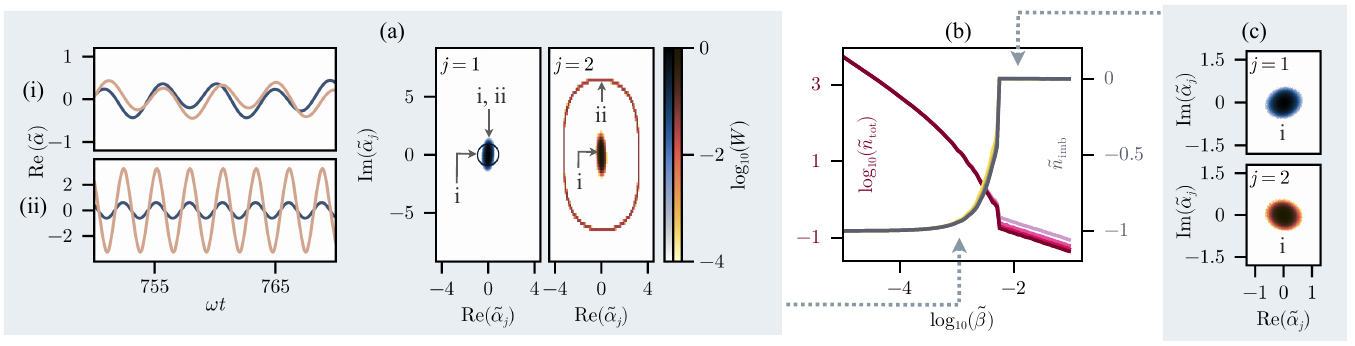}
\caption{\textit{Restoring stability via nonlinear damping.}
(a, left) At nonlinear damping strength $\tilde{\beta}=1.7\times10^{-3}$, the long-time dynamics evolve into either
(i) a low-amplitude $\PT$-like limit cycle or
(ii) a high-amplitude, population-imbalanced limit cycle, for representative initial conditions $(\tilde{\alpha}_1,\tilde{\alpha}_2)=(0.5+0.2i,\,0.3+0.1i)$ and $(2.2+0.2i,\,2.0+0.2i)$, respectively.
(a, right) Long-time single-mode phase-space distributions at $\tilde{\beta}=1.7\times10^{-3}$, with gain--loss noise retained, showing probability localized around two coexisting limit-cycle attractors (marked by symbols).
(b) Macroscopic observables, namely the total rescaled intensity $\tilde n_\mathrm{tot}$ and normalized population imbalance $\tilde n_\mathrm{imb}$, as functions of the nonlinear damping strength $\tilde{\beta}$ for $\aleph=250:250:1000$; increasing $\aleph$ corresponds to decreasing gain-loss noise strength.
(c) Same as panel (a), but for $\tilde{\beta}=0.01$, where only the low-amplitude attractor remains.}
\label{fig:wigner}
\end{figure*}

\textit{Deterministic confinement}---
We now vary the initial conditions and find two distinct dynamical fates.
Trajectories initialized near the origin remain bounded, with approximately balanced time-averaged populations, whereas sufficiently large-amplitude initial states undergo transient oscillations before the gain oscillator grows without bound.
Physically, initial population imbalances induce a differential Duffing renormalization of oscillator frequencies. 
This induced frequency mismatch dynamically decouples the oscillators, driving the gain resonator to infinity.
Bounded dynamics are therefore restricted to a finite region of phase space, see Fig.~\ref{fig:traj_bound}(a).
At exact gain-loss balance and $\beta=0$, the deterministic flow preserves phase-space volume: the Hamiltonian contribution is divergence-free, while the loss and gain contributions to the phase-space divergence cancel, $-\gamma+\gamma=0$.
This precludes asymptotically attracting sets with finite-volume basins, so the bounded region provides no restoring drift against fluctuations.
Its boundary is a global property of the nonlinear flow and cannot be determined from the local linear stability around the origin.
Nevertheless, its characteristic squared-amplitude scale varies as $\propto 1/U$~\cite{Supp}.
The bounded region therefore expands without limit as $U\to0$, recovering globally bounded deterministic dynamics in the linear limit.
Thus, the linear $\PT$ spectrum does not by itself determine global finite-amplitude stability.


Reservoir fluctuations fundamentally alter this deterministic picture.
Because the bounded sector is nonattracting, trajectories experience no restoring drift toward a stable set.
Instead, continuous fluctuations drive stochastic wandering through the bounded region until a trajectory reaches the runaway sector.
Escape is therefore a first-passage process rather than Kramers-type activation out of an attracting basin.
We simulate this dynamics within TWA using $N_\mathrm{traj}=10^4$ trajectories and define the escape time $t_\mathrm{esc}$ as the first time either oscillator exceeds the threshold $|\tilde{\alpha}_j|^2=100$.
The resulting first-passage times form a broad, nonexponential distribution that is well described empirically by a Gamma law, see Fig.~\ref{fig:traj_bound}(b).
At $\aleph=1000$, we obtain a mean escape time $\bar{t}_\mathrm{esc}=2.6\times10^4$ with relative width $\Delta t_\mathrm{esc}/\bar{t}_\mathrm{esc}=0.39$.
Increasing the scaling parameter $\aleph$ reduces the noise amplitude as $\aleph^{-1/2}$, or equivalently its variance as $\aleph^{-1}$, see Eq.~\eqref{eq: scaled noise}, and thereby prolongs the mean survival time.
Over the explored range, $\bar{t}_\mathrm{esc}$ is well described by the effective power law $\bar{t}_\mathrm{esc}\propto\aleph^{0.59}$, in contrast to the exponential-in-$\aleph$ dependence characteristic of activated escape from an attracting basin, see Fig.~\ref{fig:traj_bound}(c).
Thus, reducing the noise variance by a factor of $20$ increases the mean lifetime by only a factor of $\simeq6$.
This algebraic scaling highlights the fragility of nonattracting deterministic confinement under finite fluctuations.
For any finite noise strength studied here, the deterministically bounded sector therefore acquires a finite first-passage lifetime, while the noiseless deterministic dynamics are recovered as $\aleph\to\infty$.
Absent an additional restoring mechanism, apparently stable dynamics in the linearly $\PT$-unbroken parameter regime can consequently persist only over observation times shorter than the characteristic escape time.

\textit{Restoring stability}---
We now examine the dynamics in the presence of finite two-photon loss ($\beta\neq0$) on the gain oscillator.
This nonlinear damping explicitly breaks $\PT$ symmetry and, unlike the balanced linear gain--loss dynamics, introduces amplitude-dependent phase-space contraction.
Consequently, trajectories initialized within the previously identified bounded region evolve onto a low-amplitude $\PT$-like limit cycle, while initial conditions that previously underwent runaway converge onto a distinct high-amplitude attracting cycle, see Fig.~\ref{fig:wigner}(i,ii).
At large amplitudes, the competition between linear gain and nonlinear loss produces van der Pol-like self-saturation~\cite{Lee2013,BenArosh2021}, arresting runaway dynamics and sustaining collective oscillations of the coupled dimer.
With genuine restoring attraction present, the retained gain-loss fluctuations no longer drive irreversible first-passage escape over the simulated time window.
Instead, the long-time stochastic dynamics approach a bimodal phase-space distribution localized around the two coexisting limit-cycle attractors, see Fig.~\ref{fig:wigner}(a).

We next characterize the dependence on the nonlinear damping strength $\tilde{\beta}$.
We use the total rescaled intensity
$\tilde{n}_\mathrm{tot}=\langle|\tilde{\alpha}_1|^2+|\tilde{\alpha}_2|^2\rangle$
and the normalized population imbalance
$\tilde{n}_\mathrm{imb}
=\langle|\tilde{\alpha}_1|^2-|\tilde{\alpha}_2|^2\rangle/\tilde{n}_\mathrm{tot}$
as macroscopic diagnostics of the long-time dynamics.
Increasing $\tilde{\beta}$ suppresses the total rescaled intensity and drives the populations toward balance, see Fig.~\ref{fig:wigner}(b), since stronger nonlinear loss limits the intensity that the gain mode can sustain.
At an intermediate damping strength, both observables change discontinuously, signaling a sharp reorganization of the long-time dynamics.
This change coincides with the disappearance of the high-amplitude attracting cycle, leaving only the low-amplitude $\PT$-like attractor, see Fig.~\ref{fig:wigner}(c).
Prior to this reorganization, the long-time observables depend only weakly on the noise strength, consistent with the restoring attraction generated by nonlinear damping.
At larger $\tilde{\beta}$, stronger retained gain-loss fluctuations broaden the remaining long-time stochastic distribution and increase the mean intensity, while runaway remains suppressed over the simulated time window.

\textit{Discussion and Outlook}---
Our results separate spectral $\PT$ stability, deterministic confinement, and stochastic stability in a canonical nonlinear gain--loss dimer.
Hamiltonian Duffing nonlinearity restricts bounded dynamics to a finite but nonattracting region of phase space, so continuous reservoir fluctuations convert deterministic confinement into a finite first-passage lifetime.
Two-photon loss explicitly breaks $\PT$ symmetry, yet the resulting nonlinear damping generates genuine phase-space attraction that stabilizes limit-cycle dynamics against the same fluctuations.
Spectral gain-loss balance alone therefore does not ensure long-time stochastic stability.
Apparently stable $\PT$-like dynamics in active nonlinear platforms may instead reflect quasi-$\PT$ architectures built on passive loss imbalance without true gain, finite-time operation within a nonattracting bounded region, or restoring mechanisms such as nonlinear damping, gain saturation~\cite{Kepesidis2016}, or engineered Hamiltonian feedback~\cite{Pocklington2024}.
Because such mechanisms are often intrinsic rather than deliberately engineered, the resulting noise-resilient limit cycles are readily attributed to the $\PT$-unbroken phase, where the observed stability is real but reflects the resilience of a dissipative attractor rather than symmetry-protected linear oscillation.
More broadly, our results identify the global phase-space structure, and in particular the presence or absence of restoring attraction, as the relevant criterion for assessing long-time stability beyond the linear $\PT$ spectrum.

\textit{Acknowledgment}---
We are grateful for fruitful discussions with A.~Eichler and A.~Mikheev.
We acknowledge funding from the Deutsche Forschungsgemeinschaft (DFG) via project numbers 449653034 (Heisenberg), 425217212 (SFB1432), 521530974 (FOR5688), 545605411 (ANR), as well as from the Swiss National Science Foundation (SNSF) through the Sinergia Grant No.~CRSII5\_206008/1.
\vspace{-0.03cm}
\bibliography{PT_dimer_main}

@misc{Supp,
  author = {Tuquero, Richelle Jade L. and Seibold, Kilian and Zilberberg, Oded},
  title  = {Supplementary Material for To {$\mathcal{PT}$} or not to {$\mathcal{PT}$}},
  year   = {2026},
  note   = {Supplementary material}
}

@article{Bender1998,
  title = {Real Spectra in Non-{Hermitian} {Hamiltonians} Having {$\mathcal{PT}$} Symmetry},
  author = {Bender, Carl M. and Boettcher, Stefan},
  journal = {Phys. Rev. Lett.},
  volume = {80},
  issue = {24},
  pages = {5243--5246},
  year = {1998},
  month = {Jun},
  publisher = {American Physical Society},
  doi = {10.1103/PhysRevLett.80.5243},
  url = {https://link.aps.org/doi/10.1103/PhysRevLett.80.5243}
}

@article{Bender2007,
  title = {Making sense of non-{Hermitian} {Hamiltonians}},
  author = {Bender, Carl M.},
  journal = {Rep. Prog. Phys.},
  volume = {70},
  pages = {947--1018},
  year = {2007},
  doi = {10.1088/0034-4885/70/6/R03}
}

@article{bender2024pt,
  title = {{$\mathcal{PT}$}-symmetric quantum mechanics},
  author = {Bender, Carl M and Hook, Daniel W},
  journal = {Rev. Mod. Phys.},
  volume = {96},
  number = {4},
  pages = {045002},
  year = {2024},
  publisher = {American Physical Society},
  url = {https://link.aps.org/doi/10.1103/RevModPhys.96.045002}
}

@article{ElGanainy2018,
  title = {Non-{Hermitian} physics and {$\mathcal{PT}$} symmetry},
  author = {El-Ganainy, Ramy and Makris, Konstantinos G. and Khajavikhan, Mercedeh and Musslimani, Ziad H. and Rotter, Stefan and Christodoulides, Demetrios N.},
  journal = {Nature Physics},
  volume = {14},
  pages = {11--19},
  year = {2018},
  doi = {10.1038/nphys4323}
}

@article{Ozdemir2019,
  title = {Parity-time symmetry and exceptional points in photonics},
  author = {{\"O}zdemir, {\c S}. K. and Rotter, Stefan and Nori, Franco and Yang, Lan},
  journal = {Nature Materials},
  volume = {18},
  pages = {783--798},
  year = {2019},
  doi = {10.1038/s41563-019-0304-9}
}

@article{Heiss2012,
  title = {The physics of exceptional points},
  author = {Heiss, W. D.},
  journal = {J. Phys. A: Math. Theor.},
  volume = {45},
  pages = {444016},
  year = {2012},
  doi = {10.1088/1751-8113/45/44/444016}
}

@article{Ruter2010,
  title = {Observation of parity--time symmetry in optics},
  author = {R{\"u}ter, Christian E. and Makris, Konstantinos G. and El-Ganainy, Ramy and Christodoulides, Demetrios N. and Segev, Mordechai and Kip, Detlef},
  journal = {Nature Physics},
  volume = {6},
  pages = {192--195},
  year = {2010},
  doi = {10.1038/nphys1515}
}

@article{Makris2008,
  author  = {Makris, K. G. and El-Ganainy, R. and Christodoulides, D. N. and Musslimani, Z. H.},
  title   = {Beam Dynamics in {$\mathcal{PT}$}-Symmetric Optical Lattices},
  journal = {Phys. Rev. Lett.},
  volume  = {100},
  pages   = {103904},
  year    = {2008},
  doi     = {10.1103/PhysRevLett.100.103904},
  url     = {https://doi.org/10.1103/PhysRevLett.100.103904}
}

@article{Bender2013,
  title = {Twofold transition in {$\mathcal{PT}$}-symmetric coupled oscillators},
  author = {Bender, Carl M. and Gianfreda, Mariagiovanna and \"Ozdemir, \ifmmode \mbox{\c{S}}\else \c{S}\fi{}ahin K. and Peng, Bo and Yang, Lan},
  journal = {Phys. Rev. A},
  volume = {88},
  issue = {6},
  pages = {062111},
  numpages = {8},
  year = {2013},
  month = {Dec},
  publisher = {American Physical Society},
  doi = {10.1103/PhysRevA.88.062111},
  url = {https://link.aps.org/doi/10.1103/PhysRevA.88.062111}
}

@article{Zhang2025MechResonators,
  title = {Nonlinear distortion of nonreciprocal transmission in parity-time symmetric silicon micromechanical resonators},
  author = {Zhang, Y. and others},
  journal = {Microsyst. Nanoeng.},
  volume = {11},
  pages = {52},
  year = {2025},
  doi = {10.1038/s41378-025-00952-0}
}

@article{Schindler2011,
  title = {{$\mathcal{PT}$}-symmetric electronics},
  author = {Schindler, Joseph and Li, Ang and Zheng, Mei-Chai and Ellis, F. M. and Kottos, Tsampikos},
  journal = {Phys. Rev. A},
  volume = {84},
  pages = {040101},
  year = {2011},
  doi = {10.1103/PhysRevA.84.040101}
}

@article{Bittner2012,
  author  = {Bittner, S. and Dietz, B. and G{\"u}nther, U. and Harney, H. L. and Miski-Oglu, M. and Richter, A. and Sch{\"a}fer, F.},
  title   = {{$\mathcal{PT}$} Symmetry and Spontaneous Symmetry Breaking in a Microwave Billiard},
  journal = {Phys. Rev. Lett.},
  volume  = {108},
  pages   = {024101},
  year    = {2012},
  doi     = {10.1103/PhysRevLett.108.024101},
  url     = {https://doi.org/10.1103/PhysRevLett.108.024101}
}

@article{Lin2011,
  author  = {Lin, Zin and Ramezani, Hamidreza and Eichelkraut, Thorsten and Kottos, Tsampikos and Cao, Hui and Christodoulides, Demetrios N.},
  title   = {Unidirectional Invisibility Induced by {$\mathcal{PT}$}-Symmetric Periodic Structures},
  journal = {Phys. Rev. Lett.},
  volume  = {106},
  pages   = {213901},
  year    = {2011},
  doi     = {10.1103/PhysRevLett.106.213901},
  url     = {https://doi.org/10.1103/PhysRevLett.106.213901}
}

@article{Feng2013,
  author  = {Feng, Liang and Xu, Yi-Lin and Fegadolli, W. S. and Lu, Ming-Hui and Oliveira, J. E. B. and Almeida, V. R. and Chen, Yan-Feng and Scherer, Axel},
  title   = {Experimental demonstration of a unidirectional reflectionless parity-time metamaterial at optical frequencies},
  journal = {Nature Materials},
  volume  = {12},
  pages   = {108--113},
  year    = {2013},
  doi     = {10.1038/nmat3495},
  url     = {https://doi.org/10.1038/nmat3495}
}

@misc{bestler2025non,
  title = {Non-{Hermitian} topology and skin modes in the continuum via parametric processes},
  author = {Bestler, Markus and Dikopoltsev, Alexander and Zilberberg, Oded},
  year = {2025},
  eprint = {2505.02776},
  archivePrefix = {arXiv},
  url = {https://arxiv.org/abs/2505.02776}
}

@misc{schneider2025ultrafast,
  title = {Ultrafast Non-{Hermitian} Skin Effect},
  author = {Schneider, Barbara and Dikopoltsev, Alexander and Bestler, Markus and T{\"a}schler, Philipp and Beck, Mattias and Burghoff, David and Zilberberg, Oded and Faist, J{\'e}rome},
  year = {2025},
  eprint = {2505.03658},
  archivePrefix = {arXiv},
  url = {https://arxiv.org/abs/2505.03658}
}

@article{Wiersig2014,
  author  = {Wiersig, Jan},
  title   = {Enhancing the Sensitivity of Frequency and Energy Splitting Detection by Using Exceptional Points: Application to Microcavity Sensors for Single-Particle Detection},
  journal = {Phys. Rev. Lett.},
  volume  = {112},
  pages   = {203901},
  year    = {2014},
  doi     = {10.1103/PhysRevLett.112.203901},
  url     = {https://doi.org/10.1103/PhysRevLett.112.203901}
}

@article{Chen2017Sensitivity,
  author  = {Chen, Weijian and {\"O}zdemir, {\c S}. Kaya and Zhao, Guangming and Wiersig, Jan and Yang, Lan},
  title   = {Exceptional Points Enhance Sensing in an Optical Microcavity},
  journal = {Nature},
  volume  = {548},
  pages   = {192--196},
  year    = {2017},
  doi     = {10.1038/nature23281},
  url     = {https://doi.org/10.1038/nature23281}
}

@article{Hodaei2017,
  author  = {Hodaei, Hossein and Hassan, Absar U. and Wittek, Sebastian and Garcia-Gracia, Hipolito and {El-Ganainy}, Ramy and Christodoulides, Demetrios N. and Khajavikhan, Mercedeh},
  title   = {Enhanced Sensitivity at Higher-Order Exceptional Points},
  journal = {Nature},
  volume  = {548},
  pages   = {187--191},
  year    = {2017},
  doi     = {10.1038/nature23280},
  url     = {https://doi.org/10.1038/nature23280}
}

@article{Zhen2015,
  author  = {Zhen, Bo and Hsu, Chia Wei and Igarashi, Yuichi and Lu, Ling and Kaminer, Ido and Bromberg, Yaron and Joannopoulos, J. D. and Solja{\v c}i{\'c}, Marin},
  title   = {Spawning rings of exceptional points out of {Dirac} cones},
  journal = {Nature},
  volume  = {525},
  pages   = {354--358},
  year    = {2015},
  doi     = {10.1038/nature14889},
  url     = {https://doi.org/10.1038/nature14889}
}

@article{Weimann2017,
  author  = {Weimann, Steffen and Kremer, Mark and Plotnik, Yonatan and Lumer, Yaakov and Nolte, Stefan and Makris, K. G. and Segev, Mordechai and Rechtsman, Mikael C. and Szameit, Alexander},
  title   = {Topologically protected bound states in photonic parity--time-symmetric crystals},
  journal = {Nature Mater.},
  volume  = {16},
  pages   = {433--438},
  year    = {2017},
  doi     = {10.1038/nmat4811},
  url     = {https://doi.org/10.1038/nmat4811}
}

@article{Zezyulin2012,
  title = {Nonlinear Modes in Finite-Dimensional {$\mathcal{PT}$}-Symmetric Systems},
  author = {Zezyulin, D. A. and Konotop, V. V.},
  journal = {Phys. Rev. Lett.},
  volume = {108},
  issue = {21},
  pages = {213906},
  numpages = {5},
  year = {2012},
  month = {May},
  publisher = {American Physical Society},
  doi = {10.1103/PhysRevLett.108.213906},
  url = {https://link.aps.org/doi/10.1103/PhysRevLett.108.213906}
}

@article{Cuevas2013,
  title = {{$\mathcal{PT}$}-symmetric dimer of coupled nonlinear oscillators},
  author = {Cuevas, Jes\'us and Kevrekidis, Panayotis G. and Saxena, Avadh and Khare, Avinash},
  journal = {Phys. Rev. A},
  volume = {88},
  issue = {3},
  pages = {032108},
  numpages = {11},
  year = {2013},
  month = {Sep},
  publisher = {American Physical Society},
  doi = {10.1103/PhysRevA.88.032108},
  url = {https://link.aps.org/doi/10.1103/PhysRevA.88.032108}
}

@article{Barashenkov2014,
  doi = {10.1088/1751-8113/47/28/282001},
  url = {https://doi.org/10.1088/1751-8113/47/28/282001},
  year = {2014},
  month = {jun},
  publisher = {IOP Publishing},
  volume = {47},
  number = {28},
  pages = {282001},
  author = {Barashenkov, I V and Gianfreda, Mariagiovanna},
  title = {An exactly solvable {$\mathcal{PT}$}-symmetric dimer from a {Hamiltonian} system of nonlinear oscillators with gain and loss},
  journal = {Journal of Physics A: Mathematical and Theoretical}
}

@article{Li2011,
  title = {{$\mathcal{PT}$}-symmetric oligomers: Analytical solutions, linear stability, and nonlinear dynamics},
  author = {Li, K. and Kevrekidis, P. G.},
  journal = {Phys. Rev. E},
  volume = {83},
  issue = {6},
  pages = {066608},
  numpages = {7},
  year = {2011},
  month = {Jun},
  publisher = {American Physical Society},
  doi = {10.1103/PhysRevE.83.066608},
  url = {https://link.aps.org/doi/10.1103/PhysRevE.83.066608}
}

@misc{khedri2022fate,
  title = {Fate of exceptional points in the presence of nonlinearities},
  author = {Khedri, Andisheh and Horn, Dominic and Zilberberg, Oded},
  year = {2022},
  eprint = {2208.11205},
  archivePrefix = {arXiv},
  url = {https://arxiv.org/abs/2208.11205}
}

@book{eichler2023classical,
  title = {Classical and quantum parametric phenomena},
  author = {Eichler, Alexander and Zilberberg, Oded},
  year = {2023},
  publisher = {Oxford University Press}
}

@article{del2024limit,
  title = {Limit cycles as stationary states of an extended harmonic balance ansatz},
  author = {del Pino, Javier and Ko{\v{s}}ata, Jan and Zilberberg, Oded},
  journal = {Phys. Rev. Res.},
  volume = {6},
  number = {3},
  pages = {033180},
  year = {2024},
  publisher = {American Physical Society},
  url = {https://link.aps.org/doi/10.1103/PhysRevResearch.6.033180}
}

@article{ameye2025parametric,
  title = {Parametric instability landscape of coupled {Kerr} parametric oscillators},
  author = {Ameye, Orjan and Eichler, Alexander and Zilberberg, Oded},
  journal = {Phys. Rev. Res.},
  volume = {7},
  number = {3},
  pages = {033204},
  year = {2025},
  publisher = {American Physical Society},
  url = {https://link.aps.org/doi/10.1103/c91r-8t3h}
}

@article{Konotop2016,
  title = {Nonlinear waves in {$\mathcal{PT}$}-symmetric systems},
  author = {Konotop, Vladimir V. and Yang, Jianke and Zezyulin, Dmitry A.},
  journal = {Rev. Mod. Phys.},
  volume = {88},
  issue = {3},
  pages = {035002},
  numpages = {59},
  year = {2016},
  month = {Jul},
  publisher = {American Physical Society},
  doi = {10.1103/RevModPhys.88.035002},
  url = {https://link.aps.org/doi/10.1103/RevModPhys.88.035002}
}

@article{Suchkov2016,
  author = {Suchkov, Sergey V. and Sukhorukov, Andrey A. and Huang, Jiahao and Dmitriev, Sergey V. and Lee, Chaohong and Kivshar, Yuri S.},
  title = {Nonlinear switching and solitons in {$\mathcal{PT}$}-symmetric photonic systems},
  journal = {Laser \& Photonics Reviews},
  volume = {10},
  number = {2},
  pages = {177--213},
  doi = {10.1002/lpor.201500227},
  url = {https://onlinelibrary.wiley.com/doi/abs/10.1002/lpor.201500227},
  year = {2016}
}

@article{Scheel2018,
  doi = {10.1209/0295-5075/122/34001},
  url = {https://doi.org/10.1209/0295-5075/122/34001},
  year = {2018},
  month = {jun},
  publisher = {EDP Sciences, IOP Publishing and Societ\`a Italiana di Fisica},
  volume = {122},
  number = {3},
  pages = {34001},
  author = {Scheel, S. and Szameit, A.},
  title = {{$\mathcal{PT}$}-symmetric photonic quantum systems with gain and loss do not exist},
  journal = {Europhysics Letters}
}

@article{Huber2020,
  author   = {Huber, Julian and Kirton, Peter and Rotter, Stefan and Rabl, Peter},
  journal  = {SciPost Physics},
  title    = {Emergence of {$\mathcal{PT}$}-symmetry breaking in open quantum systems},
  year     = {2020},
  month    = oct,
  number   = {4},
  pages    = {052},
  volume   = {9},
  doi      = {10.21468/SciPostPhys.9.4.052},
  url      = {https://scipost.org/SciPostPhys.9.4.052}
}

@article{Sukhorukov2010,
  title = {Nonlinear suppression of time reversals in {$\mathcal{PT}$}-symmetric optical couplers},
  author = {Sukhorukov, Andrey A. and Xu, Zhiyong and Kivshar, Yuri S.},
  journal = {Phys. Rev. A},
  volume = {82},
  issue = {4},
  pages = {043818},
  numpages = {5},
  year = {2010},
  month = {Oct},
  publisher = {American Physical Society},
  doi = {10.1103/PhysRevA.82.043818},
  url = {https://link.aps.org/doi/10.1103/PhysRevA.82.043818}
}

@article{Ramezani2010,
  author = {Ramezani, Hamidreza and Kottos, Tsampikos and El-Ganainy, Ramy and Christodoulides, Demetrios N.},
  title = {Unidirectional nonlinear {$\mathcal{PT}$}-symmetric optical structures},
  journal = {Phys. Rev. A},
  volume = {82},
  issue = {4},
  pages = {043803},
  numpages = {6},
  year = {2010},
  month = {Oct},
  publisher = {American Physical Society},
  doi = {10.1103/PhysRevA.82.043803},
  url = {https://link.aps.org/doi/10.1103/PhysRevA.82.043803}
}

@article{Kevrekidis2013,
  doi = {10.1088/1751-8113/46/36/365201},
  url = {https://doi.org/10.1088/1751-8113/46/36/365201},
  year = {2013},
  month = {aug},
  publisher = {IOP Publishing},
  volume = {46},
  number = {36},
  pages = {365201},
  author = {Kevrekidis, Panayotis G and Pelinovsky, Dmitry E and Tyugin, Dmitry Y},
  title = {Nonlinear dynamics in {$\mathcal{PT}$}-symmetric lattices},
  journal = {Journal of Physics A: Mathematical and Theoretical}
}

@article{Dias2014,
  author = {Dias, Jo\~{a}o-Paulo and Figueira, M\'{a}rio and Konotop, Vladimir V. and Zezyulin, Dmitry A.},
  title = {Supercritical Blowup in Coupled Parity-Time-Symmetric Nonlinear {Schr\"{o}dinger} Equations},
  journal = {Studies in Applied Mathematics},
  volume = {133},
  number = {4},
  pages = {422--440},
  doi = {10.1111/sapm.12063},
  url = {https://onlinelibrary.wiley.com/doi/abs/10.1111/sapm.12063},
  year = {2014}
}

@article{Ghosh2020,
  doi = {10.1088/1751-8121/abbc50},
  url = {https://doi.org/10.1088/1751-8121/abbc50},
  year = {2020},
  month = {nov},
  publisher = {IOP Publishing},
  volume = {53},
  number = {47},
  pages = {475202},
  author = {Ghosh, Pijush K and Roy, Puspendu},
  title = {On regular and chaotic dynamics of a non-{$\mathcal{PT}$}-symmetric {Hamiltonian} system of a coupled {Duffing} oscillator with balanced loss and gain},
  journal = {Journal of Physics A: Mathematical and Theoretical}
}

@article{Martello2023,
  title = {Coexistence of stable and unstable population dynamics in a nonlinear non-{Hermitian} mechanical dimer},
  author = {Martello, Enrico and Singhal, Yaashnaa and Gadway, Bryce and Ozawa, Tomoki and Price, Hannah M.},
  journal = {Phys. Rev. E},
  volume = {107},
  issue = {6},
  pages = {064211},
  numpages = {14},
  year = {2023},
  month = {Jun},
  publisher = {American Physical Society},
  doi = {10.1103/PhysRevE.107.064211},
  url = {https://link.aps.org/doi/10.1103/PhysRevE.107.064211}
}

@article{PopDynPRR2025,
  title = {Exceptional points and stability in nonlinear models of population dynamics having {$\mathcal{PT}$} symmetry},
  author = {Felski, Alexander and Kunst, Flore K.},
  journal = {Phys. Rev. Res.},
  volume = {7},
  issue = {1},
  pages = {013326},
  numpages = {14},
  year = {2025},
  month = {Mar},
  publisher = {American Physical Society},
  doi = {10.1103/PhysRevResearch.7.013326},
  url = {https://link.aps.org/doi/10.1103/PhysRevResearch.7.013326}
}

@misc{NonHermitianDimer2026,
  title = {Global bifurcations and basin geometry of the nonlinear non-{Hermitian} skin effect},
  author = {Lin, Heng and Qi, Yunyao and Long, Gui-Lu},
  year = {2026},
  eprint = {2602.17439},
  archivePrefix = {arXiv},
  primaryClass = {quant-ph}
}

@article{Seibold2026Flow,
  author    = {Seibold, Kilian and Villa, Greta and del Pino, Javier and Zilberberg, Oded},
  title     = {Manifestations of flow topology in a quantum driven-dissipative system},
  journal   = {Phys. Rev. Res.},
  volume    = {8},
  number    = {2},
  pages     = {023093},
  year      = {2026},
  month     = {apr},
  publisher = {American Physical Society},
  doi       = {10.1103/4qh9-j88q}
}

@article{Konotop2014,
  author = {Konotop, V. V. and Zezyulin, D. A.},
  journal = {Opt. Lett.},
  number = {5},
  pages = {1223--1226},
  publisher = {Optica Publishing Group},
  title = {Stochastic parity-time-symmetric coupler},
  volume = {39},
  month = {Mar},
  year = {2014},
  url = {https://opg.optica.org/ol/abstract.cfm?URI=ol-39-5-1223},
  doi = {10.1364/OL.39.001223}
}

@article{Nowoczyn2026Melting,
  author    = {Nowoczyn, Caroline and Mathey, Ludwig and Seibold, Kilian},
  title     = {Universal quantum melting of quasiperiodic attractors in driven-dissipative cavities},
  journal   = {Phys. Rev. A},
  volume    = {113},
  number    = {5},
  pages     = {052208},
  year      = {2026},
  month     = {may},
  publisher = {American Physical Society},
  doi       = {10.1103/pdnh-1yxs}
}

@article{Nowoczyn2026Escape,
  author        = {Nowoczyn, Caroline and Mathey, Ludwig and Seibold, Kilian},
  title         = {Phase-resolved multichannel quantum escape between limit cycles},
  journal       = {arXiv e-prints},
  year          = {2026},
  month         = {may},
  eprint        = {2605.24122},
  archivePrefix = {arXiv},
  primaryClass  = {quant-ph},
  doi           = {10.48550/arXiv.2605.24122}
}

@article{Kepesidis2016,
  doi = {10.1088/1367-2630/18/9/095003},
  url = {https://doi.org/10.1088/1367-2630/18/9/095003},
  year = {2016},
  month = {sep},
  publisher = {IOP Publishing},
  volume = {18},
  number = {9},
  pages = {095003},
  author = {Kepesidis, Kosmas V and Milburn, Thomas J and Huber, Julian and Makris, Konstantinos G and Rotter, Stefan and Rabl, Peter},
  title = {{$\mathcal{PT}$}-symmetry breaking in the steady state of microscopic gain--loss systems},
  journal = {New Journal of Physics}
}

@article{Mukhamedyanov2025,
  title = {Spontaneous {$\mathcal{PT}$}-symmetry-breaking transitions under the influence of noise in an optomechanical system},
  author = {Mukhamedyanov, A. R. and Andrianov, E. S. and Zyablovsky, A. A.},
  journal = {Phys. Rev. A},
  volume = {111},
  issue = {2},
  pages = {023519},
  numpages = {7},
  year = {2025},
  month = {Feb},
  publisher = {American Physical Society},
  doi = {10.1103/PhysRevA.111.023519},
  url = {https://link.aps.org/doi/10.1103/PhysRevA.111.023519}
}

@article{Pocklington2024,
  title = {Stability via symmetry breaking in interacting driven systems},
  author = {Pocklington, Andrew and Clerk, Aashish A.},
  journal = {Phys. Rev. B},
  volume = {109},
  issue = {5},
  pages = {054309},
  numpages = {13},
  year = {2024},
  month = {Feb},
  publisher = {American Physical Society},
  doi = {10.1103/PhysRevB.109.054309},
  url = {https://link.aps.org/doi/10.1103/PhysRevB.109.054309}
}

@article{Polkovnikov2010,
  title = {Phase space representation of quantum dynamics},
  journal = {Annals of Physics},
  volume = {325},
  number = {8},
  pages = {1790--1852},
  year = {2010},
  issn = {0003-4916},
  doi = {10.1016/j.aop.2010.02.006},
  url = {https://www.sciencedirect.com/science/article/pii/S0003491610000382},
  author = {Polkovnikov, Anatoli}
}

@article{Carusotto2013,
  title = {Quantum fluids of light},
  author = {Carusotto, I. and Ciuti, C.},
  journal = {Rev. Mod. Phys.},
  volume = {85},
  issue = {1},
  pages = {299--366},
  year = {2013},
  month = {Feb},
  publisher = {American Physical Society},
  doi = {10.1103/RevModPhys.85.299},
  url = {https://link.aps.org/doi/10.1103/RevModPhys.85.299}
}

@article{Blakie2008,
  title = {Dynamics and statistical mechanics of ultra-cold {Bose} gases using c-field techniques},
  author = {Blakie, P. B. and Bradley, A. S. and Davis, M. J. and Ballagh, R. J. and Gardiner, C. W.},
  journal = {Adv. Phys.},
  volume = {57},
  number = {5},
  pages = {363--455},
  year = {2008},
  publisher = {Taylor \& Francis},
  doi = {10.1080/00018730802564254}
}

@article{Yoneya2025,
  title = {Path-integral formulation of truncated {Wigner} approximation for bosonic {Markovian} open quantum systems},
  journal = {Annals of Physics},
  volume = {479},
  pages = {170072},
  year = {2025},
  issn = {0003-4916},
  doi = {10.1016/j.aop.2025.170072},
  url = {https://www.sciencedirect.com/science/article/pii/S0003491625001538},
  author = {Yoneya, Toma and Fujimoto, Kazuya and Kawaguchi, Yuki}
}

@article{Casteels2017,
  title = {Quantum entanglement in the spatial-symmetry-breaking phase transition of a driven-dissipative {Bose-Hubbard} dimer},
  author = {Casteels, Wim and Ciuti, Cristiano},
  journal = {Phys. Rev. A},
  volume = {95},
  issue = {1},
  pages = {013812},
  numpages = {5},
  year = {2017},
  month = {Jan},
  publisher = {American Physical Society},
  doi = {10.1103/PhysRevA.95.013812},
  url = {https://link.aps.org/doi/10.1103/PhysRevA.95.013812}
}

@article{Lee2013,
  title = {Quantum Synchronization of Quantum van der {Pol} Oscillators with Trapped Ions},
  author = {Lee, Tony E. and Sadeghpour, H. R.},
  journal = {Phys. Rev. Lett.},
  volume = {111},
  number = {23},
  pages = {234101},
  year = {2013},
  doi = {10.1103/PhysRevLett.111.234101}
}

@article{BenArosh2021,
  title = {Quantum limit cycles and the {Rayleigh} and van der {Pol} oscillators},
  author = {Ben Arosh, Lior and Cross, M. C. and Lifshitz, Ron},
  journal = {Phys. Rev. Res.},
  volume = {3},
  number = {1},
  pages = {013130},
  year = {2021},
  doi = {10.1103/PhysRevResearch.3.013130}
}
\clearpage

\onecolumngrid


\begin{center}
  \textbf{\large Supplemental Material:
  To $\mathcal{PT}$ or not to $\mathcal{PT}$: Noise-induced escape and nonlinear-damping stabilization in a parity-time dimer}

  \vspace{11pt}

  Richelle Jade L.~Tuquero, Kilian Seibold, and Oded Zilberberg

  \vspace{11pt}

  \footnotesize
  \textit{Department of Physics, University of Konstanz, 78464 Konstanz, Germany}
\end{center}

\setcounter{equation}{0}
\setcounter{figure}{0}
\setcounter{table}{0}
\setcounter{section}{0}
\setcounter{page}{1}
\setcounter{secnumdepth}{3}
\renewcommand{\theequation}{S\arabic{equation}}
\renewcommand{\thefigure}{S\arabic{figure}}
\renewcommand{\thetable}{S\arabic{table}}
\renewcommand{\thesection}{S\arabic{section}}
\renewcommand{\bibnumfmt}[1]{[S#1]}
\renewcommand{\citenumfont}[1]{S#1}

\section{Escape-time statistics}
\label{app:escape}

We integrate the truncated Wigner equations [cf.~Eq.~(4) in the main text] using initial conditions sampled from a Gaussian distribution centered at the origin. Each quadrature of $\tilde{\alpha}_{1,2}$ is drawn independently with standard deviation $0.1$, all within the deterministically bounded region for all parameters considered. The same initial distribution in rescaled phase space is used for all values of $\aleph$, such that varying $\aleph$ changes the stochastic noise strength while keeping the classical initialization fixed. Unless stated otherwise, the system parameters are those used for Fig.~2 of the main text.

For each realization, we define the escape time as the first-passage time
\begin{equation}
    t_{\mathrm{esc}}
    = \inf\left\{t:\max_{j=1,2}|\tilde{\alpha}_j(t)|^2 \geq \Lambda_{\mathrm{thr}}\right\},
    \qquad \Lambda_{\mathrm{thr}}=100\,.
    \label{eq:escape_def}
\end{equation}
Trajectories that do not reach the threshold before the maximum integration time $t_{\mathrm{max}}=10^5$ are right-censored. Such trajectories constitute at most $0.14\%$ of all realizations over the parameter range considered. Given this small fraction, we neglect the censoring correction and perform the statistical fits on the escaped subset.

\begin{figure}[b]
    \centering
    \includegraphics[width=0.9\linewidth]{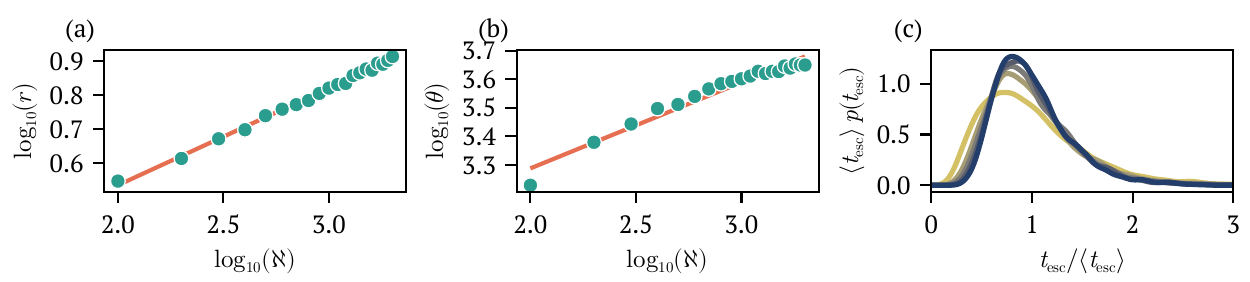}
    \caption{\textit{Scaling of escape-time statistics.} (a) Gamma shape parameter $r$ and (b) scale parameter $\theta$ versus the semiclassical scaling parameter $\aleph$, together with effective power-law fits (solid lines). (c) Escape-time probability densities expressed in rescaled variables, $t_{\mathrm{esc}}/\bar t_{\mathrm{esc}}$ and $\bar t_{\mathrm{esc}}p(t_{\mathrm{esc}})$, for varying $\aleph$; darker lines indicate larger $\aleph$.}
    \label{fig:gamma_fit}
\end{figure}

We next characterize the distribution of first-passage times. For escape from a metastable attracting state with rapid intrabasin relaxation and an approximately constant escape rate, the long-time first-passage statistics approach a single-exponential waiting-time distribution. To quantify deviations from this baseline, we empirically fit the escape times by maximum likelihood to a Gamma density,
\begin{equation}
    p(t)=\frac{t^{r-1}e^{-t/\theta}}{\Gamma(r)\theta^r}\,.
    \label{eq:gamma_density}
\end{equation}
The exponential distribution is recovered for $r=1$, whereas $r>1$ produces a distribution with a finite-time maximum. We use the Gamma law here as an empirical parametrization of the nonexponential first-passage statistics rather than as a microscopic model of the underlying stochastic dynamics.

Such nonexponential behavior is natural in the present system because the deterministically bounded region is nonattracting. Instead of relaxing toward a stable state between rare escape attempts, trajectories undergo stochastic wandering through the bounded region until they first reach the runaway sector. The resulting first-passage process therefore need not be described by a single, time-independent escape rate.

For $\aleph=1000$ and $N_{\mathrm{traj}}=10^4$, a maximum-likelihood fit to the raw, unbinned escape times yields $r=6.51$ and $\theta=4.06\times10^3$. The corresponding mean escape time and relative standard deviation are
\begin{equation}
    \bar t_{\mathrm{esc}}=r\theta=2.64\times10^4,
    \qquad
    \frac{\Delta t_{\mathrm{esc}}}{\bar t_{\mathrm{esc}}}=r^{-1/2}=0.39\,.
\end{equation}
The fitted value $r\gg1$ demonstrates a strong deviation from a single-exponential waiting-time distribution.

We repeat the analysis for $\aleph\in\{100,200,\ldots,2000\}$ using $N_{\mathrm{traj}}=10^3$ realizations for each value. Over the explored range, the fitted parameters follow effective power laws,
\begin{equation}
    r\propto\aleph^{0.285},
    \qquad
    \theta\propto\aleph^{0.301},
\end{equation}
as shown in Figs.~\ref{fig:gamma_fit}(a) and (b). Consequently,
\begin{equation}
    \bar t_{\mathrm{esc}}=r\theta\propto\aleph^{0.586},
\end{equation}
consistent with the effective scaling $\bar t_{\mathrm{esc}}\propto\aleph^{0.59}$ reported in the main text.

The fitted shape parameter increases gradually from $r\simeq3.5$ at $\aleph=100$ to $r\simeq8.2$ at $\aleph=2000$. Thus, decreasing the noise strength does not drive the statistics toward the exponential limit $r=1$ over the explored range. When expressed in units of their respective mean escape times, the distributions retain a broad, unimodal character with a maximum at finite $t_{\mathrm{esc}}/\bar t_{\mathrm{esc}}$, while narrowing systematically as $\aleph$ increases; see Fig.~\ref{fig:gamma_fit}(c). Together with the effective algebraic scaling of the mean escape time, this behavior supports the interpretation of escape as stochastic first passage through a nonattracting bounded region rather than conventional activated escape from an attracting basin.

\section{Local spectral stability and phase-space volume preservation}
\label{app:local_stability}

We first analyze the deterministic semiclassical dynamics of the Lindblad model used in the main text. Taking $\aleph\to\infty$ at fixed $\tilde U_j$ and $\tilde\beta$ eliminates the stochastic terms and the $1/\aleph$ Wigner-order corrections. We then linearize the resulting deterministic equations about the vacuum fixed point $\bm{\tilde\alpha}_{\mathrm{ss}}=\bm 0$. Defining
\begin{equation}
    \bm\delta=
    (\delta\tilde\alpha_1,\delta\tilde\alpha_1^*,
    \delta\tilde\alpha_2,\delta\tilde\alpha_2^*)^T\,,
\end{equation}
the linearized dynamics obey $\dot{\bm\delta}=\bm M_0\bm\delta$, with
\begin{equation}
\bm M_0=
\begin{pmatrix}
-i\omega_1-\frac{\gamma_1}{2} & 0 & -iJ & -iJ \\
0 & i\omega_1-\frac{\gamma_1}{2} & iJ & iJ \\
-iJ & -iJ & -i\omega_2+\frac{\gamma_2}{2} & 0 \\
iJ & iJ & 0 & i\omega_2+\frac{\gamma_2}{2}
\end{pmatrix}.
\label{eq:main_jacobian}
\end{equation}
The absence of off-diagonal elements in each conjugate pair signifies the quadrature-symmetric amplitude damping and amplification generated by the ladder-operator Lindblad channels.
The mechanically calibrated realization presented in Sec.~\ref{app:ham_cor} substitutes these entries with $\pm\gamma_j/2$, resulting in a distinct linear spectrum.
For balanced parameters $\omega_1=\omega_2\equiv\omega$ and $\gamma_1=\gamma_2\equiv\gamma$, the eigenvalues are $\lambda=\pm i\omega_\pm$, where
\begin{equation}
    \omega_\pm^2
    =\omega^2-\frac{\gamma^2}{4}
    \pm\omega\sqrt{4J^2-\gamma^2}\,,
    \label{eq:lindblad_spectrum}
\end{equation}
in agreement with Eq.~(3) of the main text. For $\gamma<\gamma_{\mathrm{EP}}=2J$, all four eigenvalues are purely imaginary. This threshold is exact for Eq.~\eqref{eq:main_jacobian}, and both $\omega_\pm^2$ remain positive throughout the interval whenever $J<\omega/2$. The vacuum is therefore locally marginal rather than asymptotically stable. At $\gamma=\gamma_{\mathrm{EP}}$ the corresponding eigenvalues coalesce at the exceptional point, while for $\gamma>\gamma_{\mathrm{EP}}$ the eigenvalues develop nonzero real parts, including exponentially growing modes. 
The two criteria described above are akin to the twofold $\mathcal{PT}$ transition seen in linked mechanical oscillators with equal loss and gain [S1-S3]. This is made exact in Sec.~\ref{app:ham_cor}.

\begin{figure}[ht]
    \centering
    \includegraphics[width=0.6\linewidth]{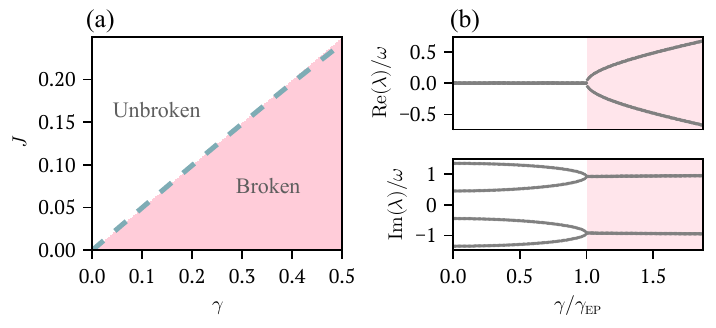}
    \caption{\textit{Local spectral stability of the main-text model.} (a) Local $\mathcal{PT}$-unbroken (white) and broken (pink) regimes of the balanced dimer, obtained by linearizing the deterministic main-text equations about the vacuum fixed point for $\tilde\beta=0$. The dashed line marks the analytical exceptional-point boundary $\gamma_{\mathrm{EP}}=2J$. (b) Real and imaginary parts of the Jacobian eigenvalues versus $\gamma$ for $\omega=0.5$ and $J=0.2$, showing their coalescence at $\gamma_{\mathrm{EP}}=0.4$.}
    \label{fig:sup_phase}
\end{figure}

In the deterministic semiclassical limit, both the Duffing term and the two-photon-loss drift are nonlinear in the amplitudes and therefore do not contribute to the Jacobian at the origin. Consequently, the nonlinear model has the same local spectral boundary as its linearized counterpart. This local result does not constrain the global finite-amplitude dynamics discussed in the main text.
At finite $\aleph$, the $1/\aleph$ Wigner correction to the two-photon-loss channel adds $+\tilde\beta/\aleph$ to each entry of the lower-right block, making the vacuum weakly repelling rather than marginal for $\tilde\beta\neq0$.

To make the phase-space contraction explicit, write
$\tilde\alpha_j=x_j+i y_j$ and define the deterministic real phase-space vector
\begin{equation}
    \bm X=(x_1,y_1,x_2,y_2)^T,
    \qquad
    \dot{\bm X}=\bm F(\bm X)\,.
\end{equation}
For $\tilde\beta=0$, the corresponding vector field is
\begin{equation}
\bm F=
\begin{pmatrix}
\omega_1y_1-\frac{\gamma_1}{2}x_1\\
-\omega_1x_1-2Jx_2-\frac{8\tilde U_1}{3}x_1^3-\frac{\gamma_1}{2}y_1\\
\omega_2y_2+\frac{\gamma_2}{2}x_2\\
-\omega_2x_2-2Jx_1-\frac{8\tilde U_2}{3}x_2^3+\frac{\gamma_2}{2}y_2
\end{pmatrix}.
\label{eq:phase_space_vector}
\end{equation}
The Hamiltonian contribution is divergence-free, while the linear loss and gain channels contribute $-\gamma_1$ and $+\gamma_2$, respectively. The Duffing terms enter only $\dot y_j$ and depend only on $x_j$, so they drop out of the divergence identically. Hence, independently of the amplitudes and of the Duffing strengths $\tilde U_j$,
\begin{equation}
    \nabla\!\cdot\bm F=-\gamma_1+\gamma_2\,.
    \label{eq:phase_space_divergence}
\end{equation}
At exact gain--loss balance, $\gamma_1=\gamma_2\equiv\gamma$, the deterministic flow is therefore phase-space-volume preserving,
\begin{equation}
    \nabla\!\cdot\bm F=0\,.
\end{equation}
In particular, the bounded sector cannot contain an asymptotically attracting set with a basin of nonzero phase-space volume. This is the origin of the distinction between local spectral boundedness and genuine restoring stability emphasized in the main text.

\section{Classical phase-space correspondence and Hamiltonian correction}
\label{app:ham_cor}

The main text formulates the active dimer using standard ladder-operator Lindblad gain and loss channels [cf.~Eqs.~(1) and (2) of the main text]. Here we establish the deterministic semiclassical correspondence between the \emph{linear} gain--loss sector of that formulation and the conventional mechanical description in which gain and loss enter as velocity-dependent forces. A standard ladder-operator Lindblad channel produces quadrature-symmetric amplitude damping or amplification and therefore does not, by itself, reproduce the conventional viscous mechanical equation in the $x$--$p$ representation. The required linear correspondence can be obtained by adding a squeezing-like Hamiltonian counterterm. Because this term changes the quantum generator, the construction below is an auxiliary mechanically calibrated realization; the spectral and stochastic results in the main text refer to the original Lindblad model, that is, to Eq.~\eqref{eq:main_jacobian} with $\hat\chi_\gamma=0$.

For equal effective masses $m$, we define physical position and momentum operators
\begin{equation}
    \hat x_j=\sqrt{\frac{\hbar}{2m\omega_j}}\,(\hat a_j+\hat a_j^\dagger),
    \qquad
    \hat p_j=-i\sqrt{\frac{\hbar m\omega_j}{2}}\,(\hat a_j-\hat a_j^\dagger).
    \label{eq:canonical_map}
\end{equation}
For the Duffing dimer without nonlinear damping, the corresponding phenomenological mechanical equations are
\begin{equation}
\label{eom:xy_linear_damping}
\begin{aligned}
\ddot{x}_1 + \omega_1^2 x_1 + \gamma_1 \dot{x}_1 + k x_2 + g_1 x_1^3  &= 0\,, \\
\ddot{x}_2 + \omega_2^2 x_2 - \gamma_2 \dot{x}_2 + k x_1 + g_2 x_2^3 &= 0\,,
\end{aligned}
\end{equation}
as in standard formulations of $\mathcal{PT}$-symmetric coupled mechanical oscillators [S1-S3]. With the convention in Eq.~\eqref{eq:canonical_map}, the conservative parameters of the main-text Hamiltonian are related to the mechanical coefficients by
\begin{equation}
    k=2J\sqrt{\omega_1\omega_2},
    \qquad
    g_j=\frac{4m\omega_j^2}{3\hbar}\,U_j\,.
    \label{eq:parameter_map}
\end{equation}

To see why a Hamiltonian correction is needed, consider first a single oscillator with a standard loss channel $\gamma\mathcal D[\hat a]\hat\rho$. Without any correction, its deterministic first moments obey
\begin{equation}
    \dot x=\frac{p}{m}-\frac{\gamma}{2}x,
    \qquad
    \dot p=-m\omega^2x-\frac{\gamma}{2}p,
\end{equation}
and therefore
\begin{equation}
    \ddot x+\gamma\dot x+\left(\omega^2+\frac{\gamma^2}{4}\right)x=0\,.
\end{equation}
This is a valid quantum-optical amplitude-damping model, but it is not identical to the conventional viscous mechanical equation with damping acting only through the velocity. The correspondence is restored by adding a squeezing-like Hamiltonian term. Writing
$\hat H=\hat H_0+\hat\chi_\gamma$, where $\hat H_0$ is the Hamiltonian of Eq.~(2) in the main text, we take
\begin{equation}
    \frac{\hat\chi_\gamma}{\hbar}
    =
    i\sum_{j=1,2}(-1)^j\frac{\gamma_j}{4}
    \left[\hat a_j^2-(\hat a_j^\dagger)^2\right].
    \label{eq:chi_gamma}
\end{equation}
This is the standard squeezing-like Hamiltonian counterterm associated with converting quadrature-symmetric ladder-operator damping into the conventional mechanical viscous form [S4-S6]. The factor $(-1)^j$ supplies the opposite signs required for loss on oscillator~1 and gain on oscillator~2. For a single loss channel, the corrected deterministic first moments become
\begin{equation}
    \dot x=\frac{p}{m},
    \qquad
    \dot p=-m\omega^2x-\gamma p,
\end{equation}
which yields $\ddot x+\gamma\dot x+\omega^2x=0$; the sign of the viscous term is reversed for gain.

The counterterm modifies the linear spectrum, and it is useful to quantify the difference.
At the semiclassical level, $\hat\chi_\gamma$ contributes $+\frac{\gamma_j}{2}\tilde\alpha_j^*$ to $\dot{\tilde\alpha}_j$ for loss, so the damping acts as $-\frac{\gamma_j}{2}(\tilde\alpha_j-\tilde\alpha_j^*)$ rather than as $-\frac{\gamma_j}{2}\tilde\alpha_j$.
Using the linearization from Sec.~\ref{app:local_stability} substitutes the zero entries of Eq.~\eqref{eq:main_jacobian} with $\pm\gamma_j/2$ and yields the characteristic polynomial $\lambda^4+(2\omega^2-\gamma^2)\lambda^2+\omega^2(\omega^2-4J^2)=0$, thus resulting in
\begin{equation}
    \left.\omega_\pm^2\right|_{\mathrm{mech}}
    =\omega^2-\frac{\gamma^2}{2}
    \pm\frac{1}{2}\sqrt{\gamma^4-4\omega^2\gamma^2+4k^2}.
    \label{eq:mech_spectrum}
\end{equation}
This is precisely the dispersion relation of the balanced coupled-oscillator system of Eq.~\eqref{eom:xy_linear_damping}, with the coupling $k$ given by Eq.~\eqref{eq:parameter_map}, and it reproduces the standard $\mathcal{PT}$ coupled-oscillator result of Ref. [S3].
The mechanically calibrated representation thus aligns precisely with the traditional $\mathcal{PT}$-symmetric oscillator framework, and both transitions of that framework are recovered.
The eigenfrequencies remain real below
\begin{equation}
    \left.\gamma_{\mathrm{EP}}\right|_{\mathrm{mech}}
    =\sqrt{2\omega^2-2\omega\sqrt{\omega^2-4J^2}},
    \label{eq:mech_ep}
\end{equation}
while the second transition, at which $\omega_-^2$ changes sign, occurs at $k=\omega^2$, equivalently $J=\omega/2$.
The exceptional point reduces to the Lindblad value $\gamma_{\mathrm{EP}}=2J$ as $J/\omega\to0$.

The divergence, by contrast, is identical in the two models.
Viscous damping removes the $-\frac{\gamma_1}{2}x_1$ contribution to $\dot x_1$ and doubles the contribution to $\dot y_1$, leaving $\nabla\!\cdot\bm F=-\gamma_1+\gamma_2$ as in Eq.~\eqref{eq:phase_space_divergence}.
hence, regardless of linear damping realization, the escape mechanism in the main text lacks an attracting set at perfect gain--loss balance.

The nonlinear two-photon-loss channel used in the main text is a distinct dissipative process and should not be identified with the phenomenological mechanical force $x^2\dot x$. To see this, consider the leading deterministic semiclassical contribution of $\beta\mathcal D[\hat a^2]\hat\rho$ to a single mode,
\begin{equation}
    \left.\dot\alpha\right|_\beta=-\beta|\alpha|^2\alpha\,.
    \label{eq:two_photon_drift_alpha}
\end{equation}
Writing $\alpha=(q+ip)/\sqrt2$ gives
\begin{equation}
    \left.\dot q\right|_\beta
    =-\frac{\beta}{2}(q^2+p^2)q,
    \qquad
    \left.\dot p\right|_\beta
    =-\frac{\beta}{2}(q^2+p^2)p,
    \label{eq:two_photon_drift_qp}
\end{equation}
with phase-space contraction rate
\begin{equation}
    \nabla_{q,p}\!\cdot\bm F_\beta
    =-2\beta(q^2+p^2)\,.
    \label{eq:two_photon_divergence}
\end{equation}
A Hermitian Hamiltonian counterterm generates divergence-free canonical flow and can therefore redistribute, but cannot alter, this contraction rate. Consequently, the two-photon Lindblad channel cannot be transformed exactly into a purely mechanical nonlinear viscous force proportional to $x^2\dot x$ by adding a Hermitian Hamiltonian counterterm alone. Such a force may be used as a separate phenomenological model of amplitude-dependent mechanical damping, but it is not identical to the two-photon-loss channel employed in the main text.

At finite $\aleph$, the two-photon-loss channel also generates multiplicative fluctuations [S7]. In the usual It\^o form of the TWA, the leading semiclassical contribution to the stochastic equation for $i\partial_t\tilde\alpha_2$ has amplitude
\begin{equation}
    \sqrt{\frac{2\tilde\beta}{\aleph}|\tilde\alpha_2|^2}\,\xi_3(t)\,,
    \label{eq:two_photon_noise}
\end{equation}
where $\xi_3$ is an independent complex Gaussian white-noise process satisfying
\begin{equation}
    \langle\xi_3^*(t)\xi_3(t')\rangle=\delta(t-t'),
    \qquad
    \langle\xi_3(t)\xi_3(t')\rangle=0\,.
\end{equation}
The corresponding noise amplitude is proportional to $|\tilde\alpha_2|$, while its variance is proportional to the active-mode population $|\tilde\alpha_2|^2$. The $1/\aleph$ deterministic Wigner correction associated with the same channel is retained explicitly in Eq.~(4) of the main text. In the simulations of the main text, we deliberately omit the multiplicative contribution in Eq.~\eqref{eq:two_photon_noise} in order to isolate the stabilizing effect of the nonlinear dissipative drift while retaining the fluctuations associated with the linear gain and loss channels. We therefore do not assume that the omitted contribution is parametrically negligible on the high-amplitude branch.

\section{Exact scaling of the bounded--runaway boundary}
\label{app:boundary_scaling}

The dependence of the deterministically bounded region on the Duffing strength can be obtained directly from a similarity transformation of the main-text equations, without introducing an empirical boundary ansatz. We set $\tilde\beta=0$, take equal Duffing nonlinearities $\tilde U_1=\tilde U_2\equiv\tilde U>0$, and restrict to the weak-coupling $\mathcal{PT}$-unbroken regime considered in the main text, $J<\omega/2$ and $\gamma<\gamma_{\mathrm{EP}}=2J$.

Introducing the dimensionless time and amplitudes
\begin{equation}
    \tau=\omega t,
    \qquad
    z_j=\sqrt{\frac{\tilde U}{\omega}}\,\tilde\alpha_j\,,
    \label{eq:z_scaling}
\end{equation}
the deterministic main-text equations become
\begin{equation}
    i\partial_\tau z_j
    =
    z_j
    +\frac{J}{\omega}(z_k+z_k^*)
    +\frac{1}{3}(z_j+z_j^*)^3
    +\frac{i}{2}s_j\frac{\gamma}{\omega}z_j,
    \qquad
    k=3-j,
    \qquad
    s_j=(-1)^j\,.
    \label{eq:scaled_boundary_eom}
\end{equation}
The Duffing strength has disappeared completely. Therefore, at fixed $J/\omega$ and $\gamma/\omega$, the entire bounded--runaway boundary is independent of $\tilde U$ in the $z_j$ variables. Because the damping term is linear in the amplitudes in either realization, this conclusion is unaffected by the counterterm of Sec.~\ref{app:ham_cor}.

Let $\mathcal B_{\tilde U}$ denote the bounded--runaway boundary in the four-dimensional space of $(\tilde\alpha_1,\tilde\alpha_2)$. For any reference value $\tilde U_0>0$, the similarity transformation implies the exact set relation
\begin{equation}
    \mathcal B_{\tilde U}
    =
    \sqrt{\frac{\tilde U_0}{\tilde U}}\,
    \mathcal B_{\tilde U_0}\,.
    \label{eq:boundary_set_scaling}
\end{equation}
Thus every linear phase-space scale of the boundary varies as $\tilde U^{-1/2}$. In particular, defining the homogeneous quadratic amplitude
\begin{equation}
    \tilde R^2
    =
    |\tilde\alpha_1|^2+|\tilde\alpha_2|^2,
\end{equation}
any characteristic boundary value obeys
\begin{equation}
    \tilde R_{\mathrm b}^2\propto\tilde U^{-1}\,.
    \label{eq:boundary_scaling_tilde}
\end{equation}
Since $\tilde\alpha_j=\alpha_j/\sqrt\aleph$ and $\tilde U=\aleph U$, the corresponding unscaled quadratic amplitude
\begin{equation}
    R^2=|\alpha_1|^2+|\alpha_2|^2
\end{equation}
satisfies
\begin{equation}
    R_{\mathrm b}^2\propto U^{-1}\,.
    \label{eq:boundary_scaling_unscaled}
\end{equation}
The bounded region therefore expands without limit as the Duffing nonlinearity is removed, recovering the globally bounded deterministic dynamics of the linear $\mathcal{PT}$-unbroken model in the parameter regime considered here.

\bigskip
\noindent\rule{\linewidth}{0.4pt}
\smallskip

\begin{list}{}
{
 \itemindent -0.3em
 \labelsep 0.4em
 \labelwidth 1.8em
 \itemsep 4pt
 \parsep 0pt}
\item[{[S1]}] C. M. Bender and S. Boettcher. Real spectra in non-Hermitian Hamiltonians having $\mathcal{PT}$ symmetry. Phys. Rev. Lett., 80:5243--5246, Jun 1998.
\item[{[S2]}] C. M. Bender, M. Gianfreda, and S. P. Klevansky. Systems of coupled $\mathcal{PT}$-symmetric oscillators. Phys. Rev. A, 90:022114, Aug 2014.
\item[{[S3]}] C. M. Bender, M. Gianfreda, \c{S}.K. \"Ozdemir, B. Peng, and L. Yang. Twofold transition in $\mathcal{PT}$-symmetric coupled oscillators. Phys. Rev. A, 88:062111, Dec 2013.
\item[{[S4]}] A. V. Dodonov, S. S. Mizrahi, and V. V. Dodonov. Quantum master equations from classical Lagrangians with two stochastic forces. Phys. Rev. E, 75:011132, Jan 2007.
\item[{[S5]}] S. N. A. Duffus, V. M. Dwyer, and M. J. Everitt. Open quantum systems, effective Hamiltonians, and device characterization. Phys. Rev. B, 96:134520, Oct 2017.
\item[{[S6]}] M. J. Everitt, W. J. Munro, and T. P. Spiller. Quantum-classical crossover of a field mode. Phys. Rev. A, 79:032328, Mar 2009.
\item[{[S7]}] T. Yoneya, K. Fujimoto, and Y. Kawaguchi. Path-integral formulation of truncated Wigner approximation for bosonic Markovian open quantum systems. Annals of Physics, 479:170072, 2025.
\end{list}

\end{document}